\documentclass[showpacs,preprintnumbers,amssymb]{revtex4}
\usepackage{graphicx}
\usepackage{epsfig}
\usepackage{dcolumn}
\usepackage{bm}

\def\be{\begin{equation}}
\def\ee{\end{equation}}
\def\bea{\begin{eqnarray}}
\def\eea{\end{eqnarray}}
\def\beqa{\begin{eqnarray*}}
\def\eeqa{\end{eqnarray*}}
\def\nnb{\nonumber}
\def\ie{{\em i.e.\relax\ }}

\def\slh{\!\!\!{/}}
\def\gsim{\lower0.5ex\hbox{$\:\buildrel >\over\sim\:$}}
\def\lsim{\lower0.5ex\hbox{$\:\buildrel <\over\sim\:$}}

\begin{document}

\title{$B\to X_s\gamma\gamma$ and $B_s\to\gamma\gamma$
in supersymmetry with broken R-parity}
\author{Alexander Gemintern}
\email[phcga@physics.technion.ac.il]{}
\affiliation{Technion--Israel Institute of Technology,\\
             32000 Haifa, Israel}
\date{\today}

\author{Shaouly Bar-Shalom}
\email[shaouly@physics.technion.ac.il]{}
\affiliation{Technion--Israel Institute of Technology,\\
             32000 Haifa, Israel}

\author{Gad Eilam}
\email[eilam@physics.technion.ac.il]{}
\affiliation{Technion--Israel Institute of Technology,\\
             32000 Haifa, Israel}
\date{\today}

\begin{abstract}
We examine the effects of R-parity violating (RPV)
supersymmetry on the two-photon B decays 
$B \to X_s \gamma \gamma$ and $B_s \to \gamma \gamma$. 
We find that, although there are many one-loop RPV diagrams 
that can contribute to these two-photon B decays, the RPV 
effect is dominated by a single diagram. This diagram, named 
here $\lambda$-{\it irreducible},   
has a distinct topology
which is irrelevant for the 
$b \to s \gamma$ amplitude at one-loop and has thus 
a negligible effect on the one-photon decay $B \to X_s \gamma$. 
We show that
the $\lambda$-{\it irreducible} RPV diagram can 
give $BR(B_s \to \gamma \gamma) \sim 5 \times 10^{-6}$
and $BR(B \to X_s \gamma \gamma) \sim 6 \times 10^{-7}$, which 
is about 16 and 5 times larger than the SM values, respectively.
Although the enhancement to the decay 
width of $B \to X_s \gamma \gamma$ is not that dramatic, we find 
that the energy distribution of the two photons 
is appreciably different from 
the SM, due to new threshold effects caused by the 
distinct topology of the RPV $\lambda$-{\it irreducible}
diagram.
Moreover, this diagram 
significantly changes the forward-backward asymmetry with respect 
to the softer photon in $B \to X_s \gamma \gamma$. 
Thus, the 
RPV effect in $B \to X_s \gamma \gamma$ can be  
discerned using these observables.     
\end{abstract}

\maketitle

\section{Introduction}
Supersymmetry (SUSY) is considered to be one of the most 
promising candidates 
for new physics, curing some of the shortcomings of the 
Standard Model (SM), for instance, the emergence of quadratic 
divergences in the Higgs sector \cite{SUSY}. 
However, despite its appealing theoretical 
features, so far there is no direct experimental evidence for SUSY
up to the electroweak scale. Apart from
direct production and decays of SUSY particles, indirect
probes could be employed to search for SUSY in
processes were the
SUSY particles emerge virtually either in loops or as 
tree-level mediators. In this respect, Flavor Changing 
Neutral Current (FCNC) or Lepton Flavor Violating (LFV) processes are 
a natural search ground for SUSY, since the SM contribution to 
such processes is 
either loop-suppressed (FCNC) or it is essentially absent (LFV).

If R-parity ($R_p$) is violated in the SUSY superpotential, 
then such flavor changing 
transitions can emerge from interactions of squarks or 
sleptons with fermions. 
R-parity is defined by $R_p\equiv (-1)^{3(B-L)+2s}$, 
where 
$B$ stands for 
baryon number, $L$ for lepton number, 
and $s$ is the spin of the particle.
Thus, $R_p=1$ for
all particles, while $R_P=-1$ for sparticles. 
The RPV 
terms in the superpotential that we will employ are:

\begin{equation} \label{lvio}
W_{RPV}=\frac{1}{2}\lambda_{ijk}\epsilon_{ab} 
\hat L_i^a \hat L_j^b 
\hat E_k^c +
\lambda'_{ijk}\epsilon_{ab}\hat L_i^a \hat Q_j^b \hat D_k^c~,
\end{equation}

\noindent where $\hat Q$ and $\hat L$ are SU(2) doublet quark 
and lepton supermultiplet, respectively, 
and $\hat D^c$ and $\hat E^c$ denote the SU(2) 
singlet down-type quark and lepton supermultiplet, respectively.
Also, $i,j,k$ are generation indices and  
$\lambda_{ijk}=-\lambda_{jik}$ due to 
the antisymmetric SU(2) indices $a,b$.
The above RPV operators may lead to  
some drastic changes in SUSY phenomenology. For example,  
the lightest sparticle becomes unstable and decays to 
SM particles and 
single sparticles may be produced in collider experiments.  

In this paper we examine the effects of the RPV SUSY 
sector on the two-photon $b \to s$ transition amplitude   
$b \to s \gamma \gamma$, focusing on the two decay channels 
$B_s \to \gamma \gamma$ and $B \to X_s \gamma \gamma$. 
These two processes have received considerable attention in 
the past decade or so \cite{LLY,SW,HK-92,RRS-96,RRS-97,BM-98,
Chin,HLX-03,singer}. 
Let us denote by $BR^{M}(B\to X_s\gamma\gamma)$ and 
$BR^{M}(B_s \to \gamma\gamma)$ the branching ratios calculated 
within a given model $M$. 
In the SM, the one-loop ElectroWeak (EW) diagrams give \cite{RRS-96,HK-92} 
(see also the next sections): 
$BR^{SM}(B\to X_s\gamma\gamma) \sim 
BR^{SM}(B_s \to \gamma\gamma) \sim 10^{-7}$.
The leading order QCD corrections to these decays can increase the
SM branching ratios by more than $100\%$ \cite{RRS-97,Chin}.

The effects of physics beyond the SM on 
the $b \to s \gamma \gamma$ amplitude 
have also been considered.  
In the 2-Higgs Doublet Model (2HDM) the $BR(B \to X_s \gamma\gamma)$ 
can range from $0.1 \times BR^{SM}(B \to X_s \gamma\gamma)$ to 
$10 \times BR^{SM}(B \to X_s \gamma\gamma)$ \cite{RRS-96,Chin} and 
in a four generation model 
$BR(B_s \to \gamma\gamma) \sim 10 \times BR^{SM}(B_s \to \gamma\gamma)$ 
\cite{HLX-03}. 
In addition, the $b \to s \gamma \gamma$ transition amplitude was 
investigated within R-parity conserving (RPC) SUSY in \cite{BM-98}, 
where only a subset of the RPC SUSY one-loop diagrams was included  
(i.e., diagrams with charged Higgs and chargino exchanges, 
neglecting possible flavor changing neutralino and gluino exchanges).  
With these assumptions,  
\cite{BM-98} found that, 
$BR^{RPC}(B_s \to \gamma\gamma)$ is highly correlated 
to $BR^{RPC}(B \to X_s \gamma)$ and is, therefore, 
bound to be within $\pm 30\%$
of the SM prediction, due to the constraints from the measured value 
of $BR(B \to X_s \gamma)$. 

The purpose of this work is to estimate only
the effects of the RPV sector on the $b \to s \gamma \gamma$ amplitude.
We, therefore, 
allow ourselves to disregard potential contributions 
that can change our results for 
$BR(B_s \to \gamma \gamma)$ and $BR(B \to X_s \gamma \gamma)$
by less than an order of magnitude,
such as the effects of RPC SUSY mentioned above, QCD corrections 
and possible long-distance non-perturbative effects.
The latter includes strong resonance effects such as 
$B \to X_s \eta (\eta^\prime) \to X_s \gamma \gamma$ and 
the even more significant $b \to s \eta_c \to s \gamma \gamma$ - 
estimated in \cite{RRS-97} to have a width about six times larger than the 
short distance width.   
These resonant contributions 
can, however, be essentially removed using 
appropriate kinematical cuts on the two photons invariant mass, with
little impact on the short distance width \cite{RRS-97}.   

As for the RPC SUSY contribution, one can alternatively assume 
that the RPC SUSY parameter space falls 
into a ``corner'' for which its effect on the $b \to s \gamma \gamma$ 
amplitude is much smaller than the RPV effect reported in this work.    
Besides, 
whether such a corner of the RPC SUSY parameter space is realized or not 
in nature, our results do not justify a detailed analysis which 
includes the above elements. 
Nonetheless, in order to 
appreciate the relative size of the RPV SUSY effect, 
we will include the EW SM contribution, defining the total width 
as (for each of the two photon decays):

\begin{eqnarray}
\Gamma=\Gamma_{SM}+\Gamma_{RPV}+\Gamma_{interference}~,
\end{eqnarray}   

\noindent where the pure SM and RPV contributions as well 
as their interference 
will be explicitly given.   
 
As in \cite{RRS-96,RRS-97,Chin}, the width 
$\Gamma(B\to X_s\gamma\gamma)$ will be 
approximated by the quark process $\Gamma(b\to s\gamma\gamma)$ and 
the branching ratio will be defined via:

\begin{eqnarray}
BR(B\to X_s\gamma\gamma) \equiv \frac{\Gamma(b \to s \gamma \gamma)}
{\Gamma(b \to c e \nu_e)} \times BR^{exp}(B \to X_c e \nu_e) ~,
 \label{brbtoxs}
\end{eqnarray}

\noindent where $\Gamma(b \to c e \nu_e) = 3 \times 10^{-5}$ eV 
is calculated at tree-level (also without QCD corrections) 
and we take 
$BR^{exp}(B \to X_c e \nu_e) = 0.11$ \cite{PDG}.
 
Following \cite{HK-92},
$\Gamma(B_s\to\gamma\gamma)$ will be calculated 
using the static quark approximation. The corresponding branching ratio 
will be defined as:

\begin{eqnarray}
BR(B_s\to \gamma\gamma) \equiv \frac{\Gamma(B_s \to \gamma \gamma)}
{\Gamma_{tot}(B_s)}~,
 \label{brbtogg}
\end{eqnarray}

\noindent where $\Gamma_{tot}(B_s)$ is the total $B_s$ width given by 
its lifetime $\tau(B_s)=1.46 \times 10^{-12}$ sec \cite{PDG}.
 
Furthermore, we define the ratios

\bea \label{r}
R^M_{s\gamma \gamma} \equiv \frac{BR^M(B \to X_s \gamma \gamma)}
{BR^{SM}(B \to X_s \gamma \gamma)}~,~
R^M_{\gamma \gamma} \equiv \frac{BR^M(B_s \to \gamma \gamma)}
{BR^{SM}(B_s \to \gamma \gamma)}~,~
R^M_{\gamma} \equiv \frac{BR^M(B\to X_s\gamma)}
{BR^{SM}(B\to X_s\gamma)}~, 
\eea

\noindent where $M$ denotes the model used 
for the calculation (RPV in our case). 
Thus, the more $R^M_{s \gamma \gamma}$, $R^M_{\gamma \gamma}$ 
or $R^M_{\gamma}$
become larger than 1, the more pronounced will the effects 
of new physics be
in the decays $B\to X_s \gamma\gamma$, $B_s \to \gamma \gamma$ or 
$B\to X_s\gamma$, respectively.

\begin{figure}
\includegraphics[width=10cm]{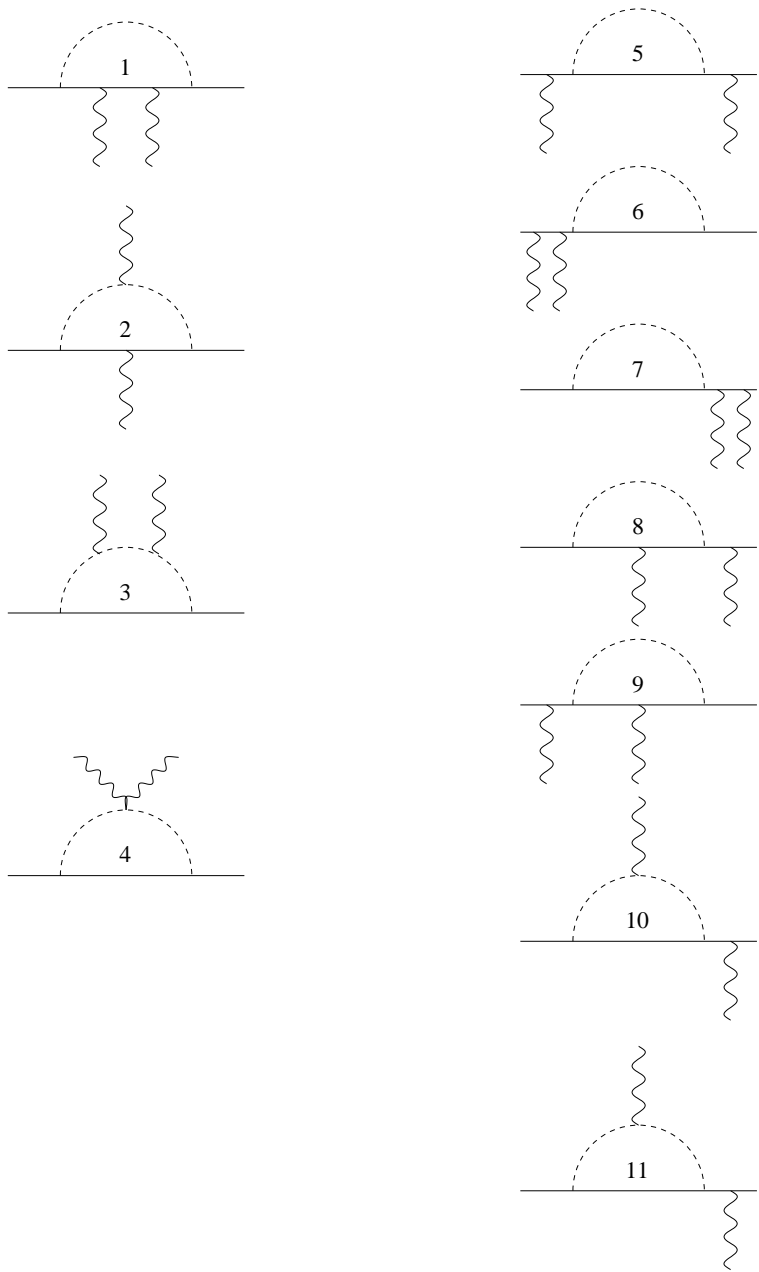}
\caption{\label{lred}
The full set of $\lambda$-{\it reducible} one-loop diagrams for 
$b\to s\gamma\gamma$. Diagrams 
1-4 are 1-particle-irreducible (1PI) and
diagrams 5-11 are 1-particle-reducible (1PR) diagrams. 
The particles in the loops may be neutrino and
$d$-type squark (then only diagrams 3-7, 10, and 11 contribute), 
sneutrino and
$d$-type quark (then only diagrams 1 and 5-9 contribute), lepton and
$u$-type squark, and slepton and $u$-type quark. All the
scalar-fermion-fermion vertices are RPV. In all the diagrams except 4,
the interchange of photons is implied.}
\end{figure}

\begin{figure}
\includegraphics[width=6cm]{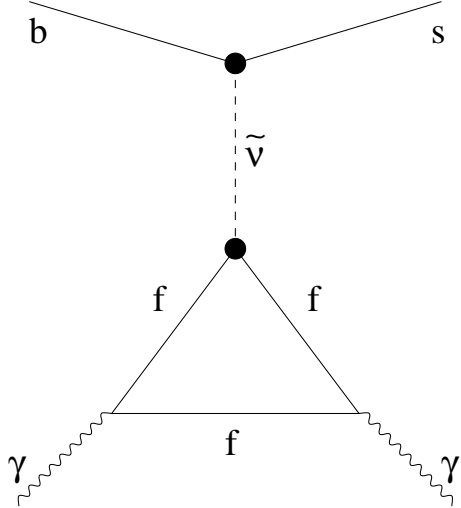}
\caption{\label{lired} A typical $\lambda$-{\it irreducible} diagram for 
$b \to s \gamma \gamma$. The full
circles denote RPV vertices. Interchange of photons is implied, and $f$,
$\tilde\nu$ stand for a fermion with weak isospin $-1/2$, sneutrino,
respectively.}
\end{figure}

Naively, one would expect the one-loop diagrams for $b \to s \gamma \gamma$ 
to be closely related to those for $b \to s \gamma$ by ``erasing'' 
one photon line. In what follows we will refer to these type 
of diagrams (shown in Fig.~\ref{lred}) as $\lambda$-{\it reducible} diagrams. 
In this case, there will be a strong correlation between 
the two and the one photon $b \to s$ decays. In other words, 
for a given new physics $M$, we expect the $\lambda$-{\it reducible} 
diagrams to yield 
$R^M_{s \gamma \gamma},~R^M_{\gamma \gamma} \sim R^M_{\gamma}$, 
and so the rather stringent 
constraint from the experimentally well measured decay $b \to s \gamma$ 
will project to the two photon decays $B \to X_s \gamma \gamma$ 
and $B_s \to \gamma \gamma$ as well. 
Indeed, such a strong correlation between the two and one photon $b \to s$ 
decays was found in the RPC SUSY case \cite{BM-98} 
and for the 2HDM \cite{Chin}. 
In contrast, as will be shown in this paper, 
in the RPV SUSY case there is a new class of diagrams 
(shown in Fig.~\ref{lired}) which are 
topologically different, thus contributing at one-loop only to 
$b \to s \gamma \gamma$ and not to $b \to s \gamma$.           
These diagrams were introduced by us \cite{GBEK} for the related 
LFV two photon decay $\mu \to e \gamma \gamma$ in \cite{GBEK}.
Following \cite{GBEK}, these diagrams will be named 
$\lambda$-{\it irreducible}, since one cannot turn them into $b\to
s \gamma$ diagrams by ``erasing'' one photon line.   
Therefore, at least in principle, in the RPV case 
the $\lambda$-{\it irreducible} 
diagrams can have a large effect in 
$B\to X_s\gamma\gamma$ or in $B_s\to\gamma\gamma$, while 
giving a much smaller (i.e., two-loop) contribution to   
$B\to X_s\gamma$. Again, such an anti-correlation can be parametrized 
by the ratios $R^{RPV}_{s \gamma \gamma}$, $R^{RPV}_{\gamma \gamma}$ 
and $R^{RPV}_{\gamma}$, since 
in this case (i.e., $\lambda$-{\it irreducible} RPV effect) 
it is possible to have 
$R^{RPV}_{s\gamma \gamma},~R^{RPV}_{\gamma \gamma}>> R^{RPV}_{\gamma}$. 

Indeed, we find that the dominant RPV effect is generated by the  
$\lambda$-{\it irreducible} diagram with a $\tau$-loop, 
which has a negligible effect 
on $b \to s \gamma$, i.e.,  giving $R^{RPV}_\gamma << 1$. 
For the case of $B\to X_s\gamma\gamma$, this diagram gives 
$R^{RPV}_{s \gamma \gamma} \sim 5$, so 
there is no dramatic change to the rate. Nonetheless, 
in spite of the rather marginal RPV effect, we find that other observables 
such as the shape of the energy distribution of the two photons 
and the value of 
a Forward-Backward-Asymmetry (FBA) (with respect to the softer 
photon), are significantly distinguishable from their SM 
counterparts in 
the presence of the 
$\lambda$-{\it irreducible} RPV effect and 
can, therefore, be used to disentangle the new physics contribution to
$B\to X_s\gamma\gamma$.  Let us also mention that we adhere to
the single  coupling scheme i.e., we consider only one pair of 
couplings at a time. 

As for the decay $B_s\to\gamma\gamma$, we find that the 
$\lambda$-{\it irreducible} diagram with a $\tau$-loop 
can enhance its branching ratio by more than an order of magnitude.
In particular, $R^{RPV}_{\gamma \gamma} \sim 16.6$, where 
the enhancement comes from the pure  
RPV contribution since the interference 
of the $\lambda$-{\it irreducible} $\tau$-loop diagram with the SM 
diagrams is much smaller.   
 
The paper is organized as follows:
In section \ref{sec:calc} we calculate the branching ratio, 
the energy distribution 
of the two photons and the FBA 
for $B \to X_s \gamma \gamma$.
In section \ref{sec:bs2gg} we calculate the branching ratio for the decay 
$B_s
 \to\gamma\gamma$ and in section \ref{sec:conc} 
we summarize our results.

\section{$B\to X_s\gamma\gamma$} \label{sec:calc}

\subsection{Calculational setup} \label{IIA}

For the calculation of the $b(p_b) \to s(p_s) \gamma(k_1) \gamma(k_2)$ decay 
rate we will employ the same cuts as in
\cite{RRS-96}. In particular, the integration domain $D$ is defined 
using the following cuts: 
\begin{enumerate}
\item The invariant mass of any pair of particles is 
constrained via: 
\bea \label{c}
(p_s+k_1)^2>cm_b^2,~(p_s+k_2)^2>cm_b^2,~(k_1+k_2)^2>cm_b^2~,
\eea
\noindent where all momenta are taken in the $b$-quark rest frame.
The "cutoff" parameter $c$
will be set to 0.01 or 0.02. 
\item The angle between any pair of the outgoing 
particles is restricted to be larger
than $20^0$. 
\item The energy of each of the photons is cut off from below at 
100 MeV, to avoid IR divergences from too soft photons in the final state.
\end{enumerate}

Thus, $\Gamma(b\to s\gamma\gamma)$ is calculated from
\bea \label{psi}
\Gamma(b\to s\gamma\gamma)=
\frac{1}{2}\int_D\frac{dE'd\omega}{64\pi^3m_b}\overline{|{\cal M}|^2},
\eea
\noindent where the factor of half is included to take 
into account the identical 
photons in the final state and $\int_D$ denotes an integration over 
the domain $D$ defined above. Also, $E'$
and $\omega$ are $s$-quark and $\gamma$ energies, respectively.
The RPV amplitudes for each diagram are defined as: 

\bea \label{uou}
{\cal M}_i=\frac{\alpha G_{\lambda}}{\sqrt{2}{\pi}}\bar{s}{\cal O}_ib~,
\eea

\noindent where $G_{\lambda}/\sqrt{2}\equiv (\lambda'\lambda')_i/(8M_i^2)$,
$(\lambda' \lambda')_i$ denotes the product of the trilinear RPV
coupling relevant for the amplitude ${\cal M}_i$ and $M_i$ is the mass of 
the corresponding sparticle. Thus, in (\ref{psi})
${\cal M}=\sum_i {\cal M}_i$ is to be understood.
 
The quark masses are taken to be: $m_t=175$ GeV, $m_b=4.5$ GeV,
$m_c=1.5$ GeV, and (as in \cite{RRS-96}) 
$m_s$ is taken to be 0.15 GeV in ${\cal M}$ and 0.45
GeV in the integration limits.

The branching ratio for the decay $B \to X_s \gamma \gamma$, subject to
the above integration domain for $b \to s \gamma \gamma$, is
then calculated using (\ref{brbtoxs}).

In addition to the rate, we will 
consider 
the energy distribution of the two outgoing
photons $1/\Gamma~d\Gamma/ d\hat{s}$, where

\bea \label{shat}
\hat{s}\equiv\frac{(k_1+k_2)^2}{m_b^2}=\frac{m_b^2+m_s^2-
2m_bE'}{m_b^2},
\eea

\noindent and the FBA
defined as follows \cite{RRS-96}:$^{[1]}$\footnotetext[1]{The FBA 
is usually defined
for processes
with distinct final particles, such as $\ell^+$ and $\ell^-$ in the decay
$b\to s\ell^+\ell^-$. 
Since in our case there are two identical particles in the final state,
the usual definition
of a FBA does not apply.}

\bea \label{eq:fba}
A_{FB}=\frac{\Gamma(\cos\theta\geq 0)
-\Gamma(\cos\theta<0)}{\Gamma(\cos\theta\geq 0)+\Gamma(\cos\theta<0)},
\eea
where $\theta$ is the angle between the $s$-quark and the softer photon.

\subsection{$B \to X_s \gamma \gamma$ in the SM}

The SM width for  $B \to X_s \gamma \gamma$ is calculated
using (\ref{brbtoxs}), with the SM amplitude given in \cite{HK-92} 
and with the set of cuts and input parameters outlined in the previous 
section. Our SM results for the branching ratio and for the 
FBA are given in Table~\ref{tabsm1}, where the contributions from the pure 
1-particle reducible (1PR) and 1-particle irreducible (1PI) diagrams, as 
well as their interference are explicitly listed.  
The energy distribution of 
the two-photons in the SM will be given and compared to the 
RPV one in the next sections.    
\begin{table}[h]
\begin{tabular}{|c|cccc|c|} \hline
& \multicolumn{4}{|c|}{$BR(B \to X_s \gamma \gamma) \times 10^7$}
& $A_{FB}$ \\
& Total  & 1PR & 1PI & Interference & \\ \hline
$c=0.01$ & 1.34 & 1.02 & 0.24 & 0.08 & 0.66 \\
$c=0.02$ & 1.18 & 0.86 & 0.24 & 0.08 & 0.63 \\ \hline
\end{tabular}
\caption{\label{tabsm1} The SM branching ratio and the FBA 
[defined in (\ref{eq:fba})], 
for the decay $B \to X_s \gamma \gamma$. The cutoff $c$ is defined in
(\ref{c}).}
\end{table}

We note that our results for the SM total branching ratio are about $15\%$ 
smaller than the results obtained in \cite{RRS-96} (recall that we are using 
the same set of cuts and inputs). This disagreement results only from the 
pure 1PR contribution, since our pure 1PI and 1PI-1PR interference 
parts are in perfect agreement with \cite{RRS-96}. 
Also, our results for the FBA in the SM agree with \cite{RRS-96} up-to
a few percent.       

\subsection{RPV couplings}

Extensive reviews on the constraints for RPV parameters can be found in
\cite{Drei, Dr-99, Bh,Barbier}. We note, however, that some of the entries
in \cite{Drei, Dr-99, Bh,Barbier} require a
renewal which we carry out for combinations of lambda's relevant 
for the $b \to s \gamma \gamma$ transition of our interest.
In particular, in what follows we will use limits obtained from  
$b\to s\ell^+\ell^-$ \cite{JKL}, from
$B\to PP$ \cite{He}, 
where $P$ is a pseudoscalar meson (this includes $B\to\pi\pi$,
$B\to K\pi$, and $B\to K\bar{K}$ decays), from
$B\to\phi K_s$ \cite{BCCK} and from $B\to\phi\phi$ and
$B\to\phi\pi$ \cite{BEY}. 
Most of these decays are generated at one-loop in the SM but at   
tree level in RPV SUSY. 
Some of the exclusive processes mentioned above 
require modeling. For example, in \cite{He} the so-called
factorization approximation was used. Therefore, some of these bounds 
are uncertain to the level of the approximation made. 
In the following, we will discuss in more detail 
the limits on the individual RPV coupling products which apply to
the specific
diagram or set of diagrams being considered.

\subsection{Contribution from $\lambda$-{\it irreducible} diagrams}

As mentioned earlier, the $\lambda$-{\it irreducible} diagrams
with the topology shown in
Fig.~\ref{lired} could, 
in principle, give a substantial contribution to $b\to s\gamma\gamma$,
almost without any effect on $b\to s\gamma$,
since they cannot be turned into $b\to s\gamma$
diagrams by removing one photon line. 
These type of diagrams were also employed in \cite{GBEK} for the LFV decay 
$\mu \to e \gamma \gamma$. Thus, for the
$\lambda$-{\it irreducible} topology, $\Gamma(b \to s \gamma \gamma)$ can
be obtained from $\Gamma(\mu \to e \gamma \gamma)$ simply by 
interchanging $m_\mu \to m_b$ and $m_e \to m_s$ and taking the
appropriate RPV couplings. We, therefore, use 
here the results of \cite{GBEK}, with the set of kinematical cuts that
defines our integration domain.

Let us first consider the case of $f=d$ in the loop of Fig.~\ref{lired}.
This diagram is proportional to the RPV
couplings $\lambda'_{i23}\lambda'_{i11}$ or
$\lambda'_{i32}\lambda'_{i11}$ (depending on the chirality 
of the $b$ and $s$-quarks), which are
constrained by the tree-level $b\to sd\bar{d}$ \cite{He} 
(note that $b \to s d \bar d$ can be obtained from
the $\lambda$-{\it irreducible} diagram by "chopping" the loop 
and turning it into a tree-level diagram).
The constraint from \cite{He} 
induces some uncertainty from the modeling of
the exclusive decay $B\to K\pi$ (when calculated from the quark-level
process $b\to sd\bar{d}$). 
Disregarding such uncertainties, 
we obtain $BR(b\to s\gamma\gamma) \sim 10^{-9}$,
\ie about two orders of magnitudes smaller than the SM prediction.

The effect of the $\lambda$-{\it irreducible} diagrams
with $f=s~or~b$ in the loop is much smaller than that with 
$f=d$, when the existing constraints on  
the relevant $\lambda' \lambda'$ product are imposed.
Besides, the couplings $\lambda' \lambda'$ relevant for $f=s~or~b$
also contribute to the $\lambda$-{\it reducible} diagrams, 
which give a larger effect for these
specific couplings (see next section).      

For the case of $f=e,\mu$, there are rather stringent constraints
coming from the inclusive $b\to sl^+l^-$
\cite{JKL}. Thus, for example, the BR for $b \to s \gamma \gamma$
calculated with the muon loop
is $\sim {\cal O}(10^{-13})$ and even smaller for the electron loop.

For the $\lambda$-{\it irreducible} diagram with the $\tau$
loop, the situation is different. 
The relevant $\lambda' \lambda$ products that can drive the $\tau$-loop
$\lambda$-{\it irreducible} diagram are:
$\lambda'_{232}\lambda_{233}$ or $\lambda'_{223}\lambda_{233}$ 
and 
$\lambda'_{123}\lambda_{133}$ or
$\lambda'_{132}\lambda_{133}$, for the $\tilde\nu_\mu$ and 
$\tilde\nu_e$ exchanges, respectively.
However, since $\lambda_{133}$ is severely
constrained by the bound on 
the electron neutrino mass \cite{GRT}, $\lambda_{133} < 0.006$, 
the contribution from the $\tilde\nu_e$ exchange is negligible. 
Moreover, since the current bounds imply that 
$\lambda'_{223}\lambda_{233} < \lambda'_{232}\lambda_{233}$ 
(see \cite{Drei,Dr-99}),
we will investigate only the effect of the $\lambda'_{232}\lambda_{233}$
coupling, for which the current limit 
is $\lambda'_{232}\lambda_{233} < 0.0234$ for a 100 GeV sparticle mass  
\cite{Drei} (note that the constraint on 
$\lambda'_{232}\lambda_{233}$ reported in 
\cite{Dr-99} is weaker).
This constraint results from a combination of bounds 
coming from $Z$-decays \cite{BES} and 
$\tau$-decays \cite{barger}. We note that there is no useful experimental
limit on $b\to s\tau^+\tau^-$ which could in principle impose 
a stronger bound on this coupling product.
Thus, taking $\lambda'_{232}\lambda_{233} =0.0234$, 
we list in Table \ref{lirwl} 
our numerical results for the maximal $\tau$-loop 
$\lambda$-{\it irreducible} contribution to
the $BR(B \to X_s \gamma\gamma)$, to the FBA $A_{FB}$ 
defined in (\ref{eq:fba}) and to
the ratio $R_{s\gamma \gamma}^{RPV}$ defined in (\ref{r}).
We see that in the $\tau$-loop case, 
the branching ratio for $B\to X_s\gamma\gamma$ can reach  
$\sim 6 \times 10^{-7}$, which is about 5 times larger than the SM
value. This enhancement arises from the pure RPV contribution since 
the interference between the RPV and the SM contributions is negligible 
in this case (i.e., less than $1\%$ 
of the total rate). 
 
\begin{figure} 
\includegraphics[width=10cm]{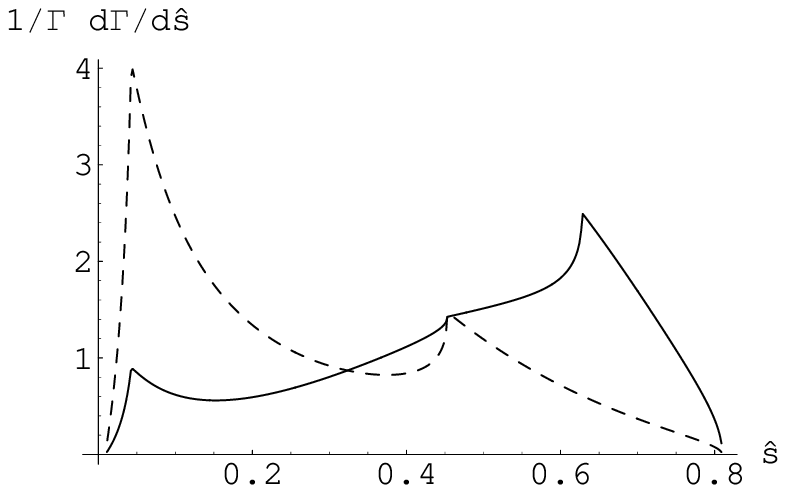}
\includegraphics[width=10cm]{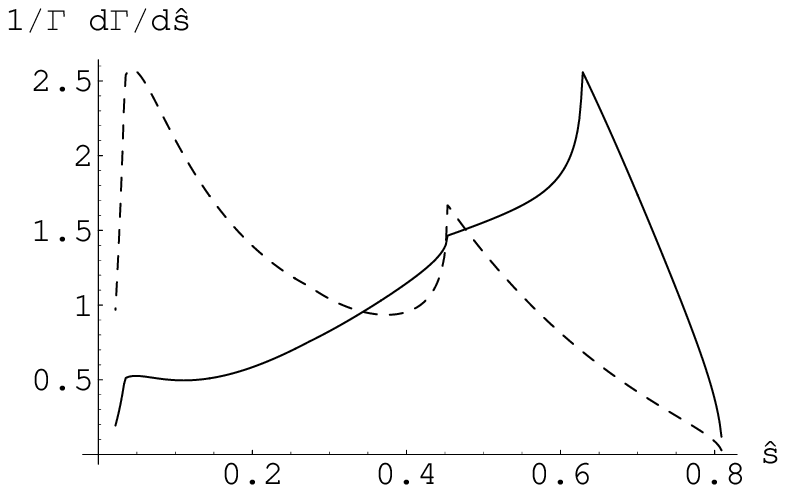}
\caption{\label{dGdE2} The two photons energy distribution 
$1/\Gamma~d\Gamma/d\hat{s}$, as a function of 
$\hat{s} \equiv (k_1+k_2)^2/m_b^2$, 
in the SM (dashed-line) and in the SM+RPV case (solid-line) 
for the $\lambda$-{\it irreducible}
diagram with the $\tau$-loop, with $\lambda'_{232}\lambda_{233}=0.0234$. 
For the upper graph $c=0.01$ while 
$c=0.02$ for the lower one. 
The low $\hat{s}$ peak
is due to bremsstrahlung and the cuts in Eqn. (\ref{c}). The other two
sharp peaks correspond to threshold openings at
$\hat{s}=4m_{\tau}^2/m_b^2$ and $\hat{s}=4m_c^2/m_b^2$.}
\end{figure}

\begin{table}[h]
\begin{tabular}{|c|c|c|cccc|c|c|} \hline
& RPV & & \multicolumn{4}{|c|}{$BR(B \to X_s \gamma \gamma) \times 10^7$}
& $A_{FB}$ & $R^{RPV}_{s \gamma \gamma}$ \\
& coupling & $(\lambda^\prime\lambda)_{max}$ \cite{Drei} & SM  & 
RPV & Interference & Total &  & \\ \hline
$c=0.01$ &$\lambda'_{232}\lambda_{233}$  & 0.0234 & 
1.34 & 4.81 & $\times$ & 6.15 & 0.46 & 4.6 \\
$c=0.02$ & -"-  &  -"- & 
1.18 & 4.81 & $\times$ & 5.99 & 0.45 & 5.1 \\ \hline
\end{tabular}
\caption{\label{lirwl} $BR(B \to X_s \gamma \gamma) \times 10^7$ for
the $\lambda$-{\it irreducible} diagram with the $\tau$-loop case. 
$R^{RPV}_{s \gamma \gamma}$ is defined in (\ref{r}) and $A_{FB}$
is defined in (\ref{eq:fba}). The sneutrino mass is assumed to be 100 GeV. 
$\times$ means that the contribution 
is of ${\cal O}(1\%)$ (or smaller) of the total 
rate. The cutoff $c$ is defined in (\ref{c}).}
\end{table}

We also find that, for the $\lambda$-{\it irreducible} diagram 
with the $\tau$-loop, 
the FBA ($A_{FB} \sim 0.45$, see Table \ref{lirwl}) as well as the photons
energy distribution 
$1/\Gamma~d\Gamma/d\hat{s}$ (shown in Fig.~\ref{dGdE2}) are 
significantly different from their values in the SM. 
Therefore, these quantities may prove useful 
for disentangling the $\lambda$-{\it irreducible} RPV effect 
in $B \to X_s \gamma \gamma$. Let us also mention that we adhere to the
one-coupling scheme, i.e., assuming one contribution of a pair of RPV
couplings at a time.
\subsection{Contribution from $\lambda$-{\it reducible} diagrams}
\begin{figure} 
\includegraphics[width=5cm]{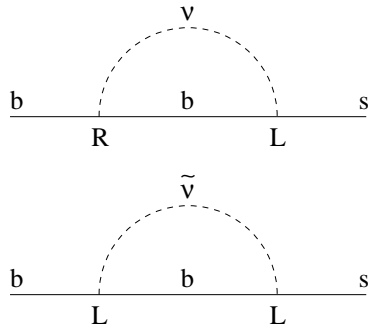}
\caption{\label{flip} The chirality non-flip (upper part) 
and chirality flip (lower part) diagrams. L and R
denote left and right chirality projectors, respectively. 
The external photons may be attached to any charged line.}
\end{figure}

The full set of one-loop 
$\lambda$-{\it reducible} RPV diagrams for $b\to s\gamma\gamma$
is given in Fig.~\ref{lred}. These diagrams can be classified 
according to the RPV coupling involved.
For example, for $\lambda'_{ij2}\lambda'_{ij3}$ all the 
$\lambda$-{\it reducible} diagrams
are relevant, while for $\lambda'_{i2k}\lambda'_{i3k}$ only the diagrams
with $d$-quark-sneutrino ($d-\tilde\nu$) or 
$d$-squark-neutrino ($\tilde{d}-\nu$) in the loop 
contribute. In addition, the 
$\lambda$-{\it reducible} diagrams
can be further subdivided (shown in Fig.~\ref{flip}), 
into "flip"-diagrams  
and "non-flip" diagrams according to whether the 
incoming $b$-quark and outgoing $s$-quark have different or the 
same chiralities, respectively. The flip diagrams are 
important only  
for $j=3$ or $k=3$, where the mass insertion for the chirality flip is $m_b$ 
(i.e., the internal quark in the loop is the $b$-quark). 
The classification of the $\lambda$-{\it reducible} diagrams 
is given in Table~\ref{class}.

\begin{table}[h]
\begin{tabular}{|c|c|c|c|c|c|c|c|c|} \hline
& \multicolumn{4}{|c|}{1PI} & \multicolumn{4}{|c|}{1PR}\\ 
loop particles& $d-\tilde{\nu}$ & $\nu-\tilde{d}$ 
& $u-\tilde{\ell}$ & $\ell-\tilde{u}$
& $d-\tilde{\nu}$ & $\nu-\tilde{d}$ 
& $u-\tilde{\ell}$ & $\ell-\tilde{u}$\\
RPV couplings& ($\lambda'_{ij3}\lambda'_{ij2},$ &
($\lambda'_{ij3}\lambda'_{ij2},$ &
($\lambda'_{ij3}\lambda'_{ij2}$) &
($\lambda'_{ij3}\lambda'_{ij2}$) &
($\lambda'_{ij3}\lambda'_{ij2},$ &
($\lambda'_{ij3}\lambda'_{ij2},$ &
($\lambda'_{ij3}\lambda'_{ij2}$) &
($\lambda'_{ij3}\lambda'_{ij2}$)\\
& $\lambda'_{i3k}\lambda'_{i2k}$) & $\lambda'_{i3k}\lambda'_{i2k}$) &&
& $\lambda'_{i3k}\lambda'_{i2k}$) & $\lambda'_{i3k}\lambda'_{i2k}$) &&\\
\hline
non-flip & 1 & 3,4 & 1-4 & 1-4 & 5-9 & 5-7,10,11 & 5-11 & 5-11\\
flip & $\surd$ & $\times$ & $\times$ & $\times$ & $\surd$ & $\times$ & $\times$ & $\times$\\
\hline
\end{tabular}
\caption{\label{class} Classification of $\lambda$-{\it reducible}
diagrams. The numbers in the "non-flip" row denote those diagrams in
Fig.~\ref{lred} which belong to the non-flip category. 
In the "flip" row $\surd$ is inserted if the relevant diagram can
have a chirality flip, and $\times$ if not. Also, 
1PI and 1PR stand for one-particle-irreducible and one-particle-reducible
diagrams, respectively.}
\end{table}

We find that the maximal effect (i.e., using the maximal allowed 
values for the corresponding RPV couplings) of any given 
non-flip diagram is always 
about two order of magnitude smaller than the contribution generated 
by the dominant flip diagrams with the $b- \tilde\nu$ loops. 
Nonetheless, for completeness, we give below the derivation
of the amplitudes for the largest non-flip contributions.

\subsubsection{Non-flip diagrams}
         
For the non-flip diagrams we define:
\bea
Q_3+\Delta Q_3=i\epsilon^{\mu\nu\xi\alpha}\gamma_{\alpha}L(k_1-k_2)_{\xi}+
i\frac{k_{1\xi}k_{2\eta}}{k_1\cdot
k_2}(\epsilon^{\mu\xi\eta\alpha}k_1^{\nu}- 
\epsilon^{\nu\xi\eta\alpha}k_2^{\mu})\gamma_{\alpha}L \label{q3},
\eea
and
\bea
W&=&-\left[\left(\frac{p_s^{\nu}}{p_s\cdot k_2}-\frac{p_b^{\nu}}{p_b\cdot
k_2}\right)\sigma(\mu,k_1)+\left(\frac{p_s^{\mu}}{p_s\cdot
k_1}-\frac{p_b^{\mu}}{p_b\cdot
k_1}\right)\sigma(\nu,k_2)\right] \nonumber \\
&&+\frac{i}{2}\left[\left(\frac{1}{p_s\cdot
k_2}-\frac{1}{p_b\cdot k_1}\right)\sigma(\nu,k_2)\sigma(\mu,k_1)+ 
\left(\frac{1}{p_s\cdot k_1}-\frac{1}{p_b\cdot k_2}\right)
\sigma(\mu,k_1)\sigma(\nu,k_2)\right] \label{W}~,
\eea

\noindent where $\sigma(\mu,k)\equiv \sigma^{\mu\rho}k_{\rho}$ and 
$\sigma^{\mu\nu}=\frac{1}{2i}\left[\gamma^{\mu},\gamma^{\nu}\right]$.
Then, using the prescription given in section \ref{IIA},
the operators ${\cal O}$ [defined in (\ref{uou})], for the 
non-flip 
diagrams that give the largest non-flip effect 
are given in Table~\ref{tab:o}, wherein the non-flip contributions 
are further classified according to the particles/sparticles 
exchanged in the loops.

\begin{table}[h]
\begin{tabular}{|c|c|} \hline
Loop particles & ${\cal O}$\\
\hline
$\tilde{b}-\nu$&
$-\frac{i}{54}W(m_bL+m_sR)$\\
$b-\tilde\nu$ (non-flip) &
$-\frac{2}{9}(Q_3+\Delta Q_3)\delta_3+\frac{i}{27}W(m_bL+m_sR)$\\
$\tilde{c}-\ell$ &
$-2(Q_3+\Delta Q_3)\delta_3+\frac{4i}{27}W(m_bL+m_sR)$\\
$c-\tilde\ell$&
$-\frac{8}{9}(Q_3+\Delta Q_3)\delta_3-\frac{7i}{54}W(m_bL+m_sR)$\\
$t-\tilde\ell$&
$-\frac{1}{3}F_2'(z^2)W(m_bL+m_sR)$,
where
$F_2'(x)=\frac{7-12x-3x^2+8x^3-6x(-2+3x)\log x}{18(x-1)^4}$
and $z=m_t/M$\\ \hline
\end{tabular}
\caption{\label{tab:o} The operators appearing in the $\lambda$-{\it
reducible} amplitudes following the definition in (\ref{uou}). 
$Q_3+\Delta Q_3$ and $W$ are given in (\ref{q3}) and (\ref{W}). 
$M$ is the sparticle mass.}
\end{table}

As mentioned above, the contribution from 
the non-flip operators in Table \ref{tab:o} 
is always subdominant compared to that coming from the 
$b-\tilde\nu$ flip diagrams.
In particular, for the non-flip case we find that the typical branching 
ratio is $BR(B \to X_s \gamma \gamma) \sim 10^{-9}$.
Let us, therefore, proceed with the $b-\tilde\nu$ flip diagrams.

\subsubsection{Flip diagrams}

The dominant flip diagrams are obtained from the 
$b-\tilde\nu$ loops and are, therefore, proportional to 
either $\lambda'_{323}\lambda'_{333}$  or
$\lambda'_{332}\lambda'_{333}$, 
depending on the chiralities of the 
incoming $b$-quark and outgoing $s$-quark.$^{[2]}$\footnotetext[2]{Note that  
the $\lambda$-{\it reducible} diagrams with RPV couplings corresponding
to $b$ and $s$-quark in the loops, contribute also to the 
$\lambda$-{\it irreducible} topology. However, for these specific couplings,
the $\lambda$-{\it reducible} effect exceeds  
the $\lambda$-{\it irreducible} one.}
The constraints on these RPV coupling products come 
from $b\to cl\nu$ and $B_s-\bar{B}_s$ mixing, see \cite{BCCK}. 
The expression for the amplitude of these flip diagrams
is, unfortunately, too long to be useful for the reader and will not 
be given here.$^{[3]}$\footnotetext[3]{The expression for the 
$b-\tilde\nu$ flip diagrams can be ordered by e-mail from S.G.}

The calculation for the flip diagrams was performed using the 
Mathematica package. The numerical results 
are presented in Table{~\ref{tab:res}, where 
all interferences were taken to be constructive, by adjusting 
the sign of the RPV couplings.

\begin{table}[h]
\begin{tabular}{|c|c|c|cccc|c|c|} \hline
& RPV & & \multicolumn{4}{|c|}{$BR(B \to X_s \gamma \gamma) \times 10^7$}
& $A_{FB}$ & $R^{RPV}_{s \gamma \gamma}$ \\
& coupling & $(\lambda^\prime\lambda^\prime)_{max}$ \cite{BCCK} & SM  & 
RPV & Interference & Total &  & \\ \hline
$c=0.01$ &$\lambda'_{323}\lambda'_{333}$  & 0.0033 & 
1.34 & 1.02 & 2.11 & 4.47 & 0.73 & 3.3 \\
         & $\lambda'_{332}\lambda'_{333}$  & 0.0025 & 
-"- & 0.58 & 0.05 & 1.97 & 0.7 & 1.5 \\ \hline
$c=0.02$ & $\lambda'_{323}\lambda'_{333}$  & 0.0033 & 
1.18 & 0.86 & 1.8 & 3.84 & 0.7 & 3.3 \\ 
         & $\lambda'_{332}\lambda'_{333}$  & 0.0025 & 
-"- & 0.49 & 0.04 & 1.71 & 0.66 & 1.5 \\ \hline
\end{tabular}
\caption{\label{tab:res} Maximal $BR(B \to X_s \gamma \gamma) \times 10^7$,
$A_{FB}$ and $R^{RPV}_{s \gamma \gamma}$, for
the $\lambda$-{\it reducible} flip diagrams with the $b - \tilde\nu$ loops. 
The cutoff $c$ is defined in (\ref{c}).}
\end{table}

\begin{figure}
\includegraphics[width=10cm]{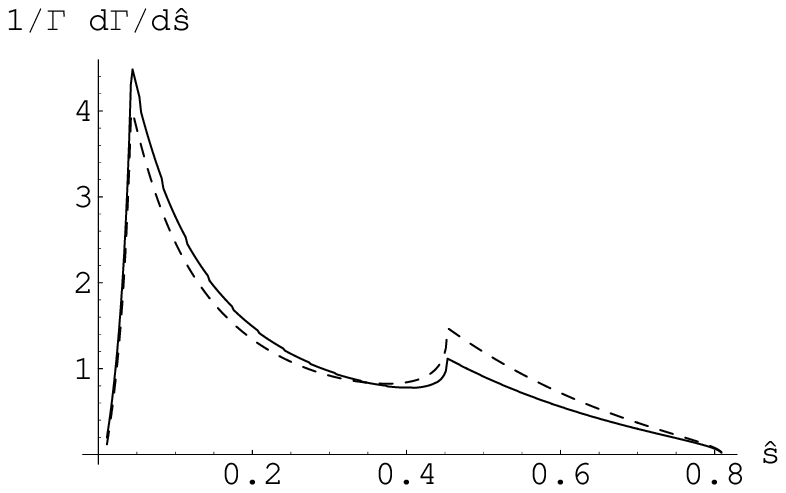}
\includegraphics[width=10cm]{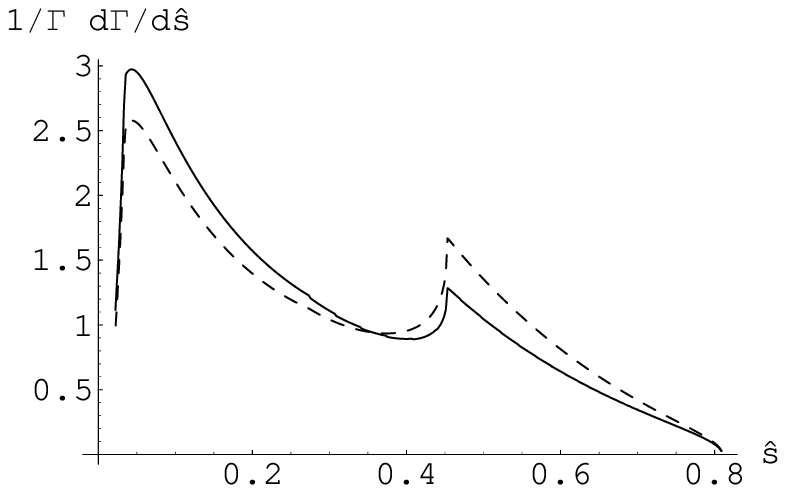}
\caption{\label{dGdE} $1/\Gamma~d\Gamma/d\hat{s}$ as a function of
$\hat{s}$
for SM (dashed) and SM+RPV (solid) for $\lambda$-{\it reducible}
diagrams with $\lambda'_{i32}\lambda'_{i33}=0.0025$. Here $c=0.01$ for
the upper graph and $c=0.02$ for the lower one.
The cutoff $c$ is defined in (\ref{c}).}
\end{figure}
 
\begin{figure}
\includegraphics[width=10cm]{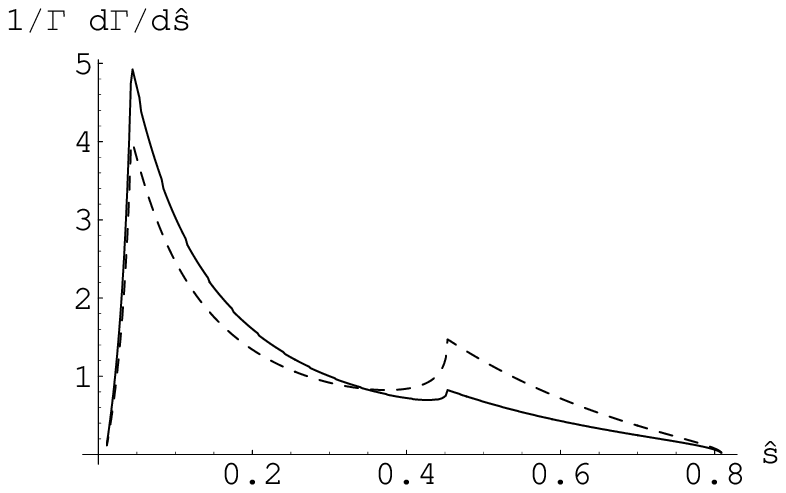}
\includegraphics[width=10cm]{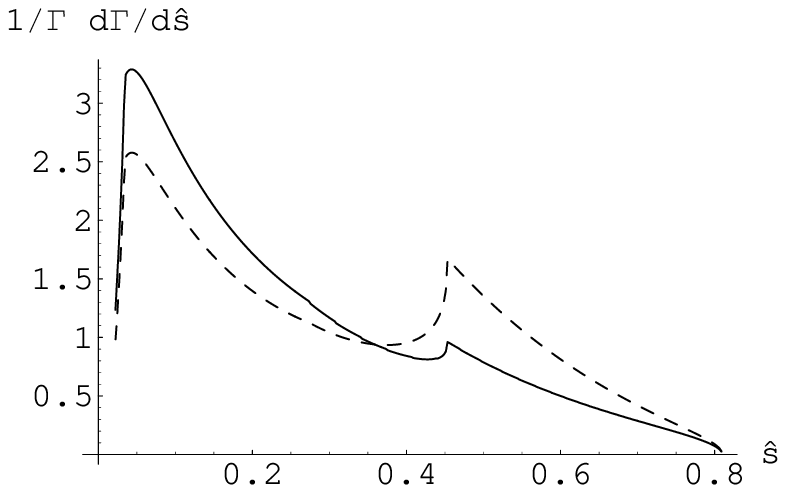}
\caption{\label{dGdE1} $1/\Gamma~d\Gamma/d\hat{s}$ as a function of
$\hat{s}$
for SM (dashed) and SM+RPV (solid) for $\lambda$-{\it reducible} diagrams
with $\lambda'_{323}\lambda'_{333}=0.0033$. Here $c=0.01$ for
the upper graph and $c=0.02$ for the lower one.
The cutoff $c$ is defined in (\ref{c}).}
\end{figure}

From Table \ref{tab:res} we see that the largest effect, which 
comes from the interference 
of the RPV $b-\tilde\nu$ flip diagrams 
($\propto \lambda'_{323}\lambda'_{333}$) with the SM ones, is not 
quantitatively much different from the pure SM prediction and is smaller 
than the $\lambda$-{\it irreducible} effect with the $\tau$-loop. 
The same conclusion holds for the overall SM+RPV FBA which remains
close to its SM value.

In Figs.~\ref{dGdE} and
\ref{dGdE1} we show the energy distribution of the two photons, 
$1/\Gamma~d\Gamma/d\hat{s}$, in the pure SM and for the SM + 
RPV flip diagrams with 
$\lambda'_{332}\lambda'_{333}=0.0025$ and 
$\lambda'_{323}\lambda'_{333}=0.0033$, respectively. 
As can be seen from these figures, the 
shape of the SM energy distribution remains almost unchanged when 
the RPV flip diagrams are included.

Therefore, to conclude this section, the largest 
RPV $\lambda$-{\it reducible} contribution to $BR(B \to X_s \gamma \gamma)$, 
which comes from flip diagrams with 
$b-\tilde\nu$ loops, is somewhat smaller than the RPV 
$\lambda$-{\it irreducible} contribution. 
Moreover, within the $\lambda$-{\it reducible} 
topology, the FBA and the photons energy 
distribution are essentially indistinguishable from the SM, whereas 
the $\lambda$-{\it irreducible} topology can significantly alter 
the value and shape of 
these quantities.

\section{$B_s\to\gamma\gamma$} \label{sec:bs2gg}
For the calculation of $\Gamma(B_s\to\gamma\gamma)$ 
we adopt the static quark approximation \cite{HK-92}, where
$m_{B_s}=m_b+m_s$ and both $b$ and $s$ have zero
3-momenta in the rest frame of the decaying $B_s$ meson, in which case the 
photons are emitted back-to-back with energies $M_{B_s}/2$. 
Since this model is non-relativistic, as in \cite{HK-92}, we 
use the constituent mass for $s$-quark $m_s=500$ MeV.
 
The current matrix element is \cite{HK-92}:
\bea \label{eq:mu5}
\langle 0|\bar{u}_s\gamma_{\mu}\gamma_5u_b|B_s\rangle=-if_{B_s}P_{\mu}~,
\eea
where $P=p_b-p_s$ is the $B_s$-meson 4-momentum and 
$f_{B_s}$ is the $B_s$-meson decay constant. 

From (\ref{eq:mu5}) one obtains 
\bea \label{eq:5}
\langle 0|\bar{u}_s\gamma_5u_b|B_s\rangle=if_{B_s}m_{B_s}~.
\eea
In addition, the fact that
\bea \label{eq:mu}
\langle 0|\bar{u}_s\gamma_{\mu}u_b|B_s\rangle=0~,
\eea
can be used to simplify the calculation for the 
$\lambda$-{\it reducible} flip diagrams. For
example, the term $\bar{u}_sL\gamma_{\mu}\gamma_{\nu}u_b$ can be
transformed to a combination
of four terms:
\bea
&&\bar{u}_sL\gamma_{\mu}\gamma_{\nu}u_b=\frac{1}{m_b}\bar{u}_sL
\gamma_{\mu}\gamma_{\nu}p\slh_b u_b=\\ &&\frac{1}{m_b} 
(g_{\mu\nu}\bar{u}_sLp\slh_b
u_b+p_{b \nu}\bar{u}_sL\gamma_{\mu}u_b-p_{b \mu}\bar{u}_sL\gamma_{\nu}u_b+
i\epsilon_{\mu\nu\rho\sigma}p^{\rho}_b\bar{u}_sL\gamma^{\sigma}u_b),
\eea
which, when put between vacuum and meson states 
as in (\ref{eq:mu5}), turns into
\bea
\bar{u}_sL\gamma_{\mu}\gamma_{\nu}u_b=\frac{if_{B_s}}{2m_b}
(g_{\mu\nu}P \cdot p_b+p_{b \nu}P_{\mu}-p_{b \mu}P_{\nu}+
i\epsilon_{\mu\nu\rho\sigma}p^{\rho}_b P^{\sigma}) \label{eq18}~.
\eea
Thus, in the static quark approximation (\ref{eq18}) amounts to
\bea
\bar{u}_sL\gamma_{\mu}\gamma_{\nu}u_b=\frac{if_{B_s}m_{B_s}}{2}g_{\mu\nu}~.
\eea
Following the above procedure, the $B_s\to\gamma\gamma$ amplitude 
can be parametrized as \cite{HK-92}:

\bea
{\cal M}_M=2 f_{B_s} \left[B_M^+
(k'_{\mu}k_{\nu}-\frac{1}{2}m_{B_s}^2g_{\mu\nu})+B_M^-i\epsilon_{\mu\nu\rho\sigma}
k^{\rho}k'^{\sigma}\right]\epsilon^{\mu}\epsilon'^{\nu} ~, 
\eea

\noindent where the subscript $M$ denotes the model used for the calculation.
The decay width for $B_s\to\gamma\gamma$ is then given by:

\bea
\Gamma^M(B_s\to\gamma\gamma) = f_{B_s}^2 \frac{m_{B_s}^3}{16 \pi} 
\left( |B_M^+|^2 + |B_M^-|^2 \right) ~. 
\eea 

\noindent In the SM 

\bea
B_{SM}^{\pm}=\frac{\alpha G_F }{\sqrt{2} \pi} A^{\pm},
\eea
 
\noindent where the form factors $A^+$ and $A^-$ are defined in \cite{HK-92}.

As in the case of $B \to X_s \gamma \gamma$, we find that the 
potentially largest RPV contribution to the width 
of $B_s \to \gamma \gamma$ comes from the 
$\lambda$-{\it irreducible} diagram with the $\tau$-loop. 
In particular, this diagram gives an enhancement to 
$\Gamma(B_s \to \gamma \gamma)$ which is about 
100 or 10 times larger than the 
one obtained from the $\lambda$-{\it reducible} non-flip or flip diagrams,
respectively.
 
The RPV form factors in the $\lambda$-{\it irreducible} case are:

\bea
B_{RPV}^{+;-}=\frac{\alpha m_{B_s}}{16\pi}
\frac{i\lambda'_{232} \lambda_{233}}{m_{\tau} M_{\tilde\nu}^2}
  \cdot f_{1/2}(x);g_{1/2}(x)~, 
\label{CLCR}
\eea
 
\noindent with

\bea\label{Functions}
f_{1/2}(x) &=& 2x\left[1+(1-x)
               \arcsin^2\left(\frac{1}{\sqrt{x}}\right)\right]\,,
               \nnb\\
g_{1/2}(x) &=& 2x\arcsin^2\left(\frac{1}{\sqrt{x}}\right) ~,
\eea

\noindent where $M_{\tilde\nu}$ 
is the sneutrino mass and for the $\tau$-loop case $x=(2m_{\tau}/m_{B_s})^2$.
  
Our numerical results for the SM and the 
$\lambda$-{\it irreducible} RPV contributions to 
$BR(B_s \to \gamma \gamma)$ [using
(\ref{brbtogg})] are summarized in Table \ref{tab:ggres}.  
The results were obtained with $m_b=4.8$ GeV, $m_s=0.5$ GeV, 
$m_{B_s}=m_b+m_s$, 
$V_{cb}=0.04$ and $f_{B_s}=200$ MeV.  
The RPV couplings were taken at their maximal allowed values and 
the sneutrino mass was set to 100 GeV.

\begin{table}[ht]
\begin{tabular}{|c|c|cccc|c|}
\multicolumn{7}{|c|}{RPV from $\lambda$-{\it irreducible} diagrams with 
the $\tau$-loop} \\ \hline
RPV & & \multicolumn{4}{|c|}{$BR(B_s \to \gamma \gamma) \times 10^7$}
& \\
coupling & $(\lambda^\prime\lambda^\prime)_{max}$ \cite{Drei} & SM  & 
RPV & Interference & Total &  $R^{RPV}_{\gamma \gamma}$ \\ \hline
$\lambda'_{232}\lambda_{233}$  & 0.0234 & 
2.74 & 38.4 & 4.41 & 45.55  & 16.62 \\ \hline
\end{tabular}
\caption{\label{tab:ggres} $BR(B_s \to \gamma \gamma) \times 10^7$
in the SM and for 
the $\lambda$-{\it irreducible} diagrams with the $\tau$-loop.  
The sneutrino mass was set to 100 GeV. See also text.}
\end{table}

We see that 
the enhancement from 
the $\lambda$-{\it irreducible} $\tau$-loop diagram
is here particularly large: $BR(B_s \to \gamma \gamma) \sim 5 \times 10^{-6}$,
about 16 times larger than the SM prediction   
(although still far below the experimental limit, 
$1.48\times 10^{-4}$ \cite{PDG}). 
We also note that our numerical result for the SM branching ratio 
agrees with 
\cite{HK-92} after substituting our $f_{B_s}$, 
$V_{cb}$ and $\tau(B_s)$ for the values used in \cite{HK-92}.

\section{Conclusions} \label{sec:conc}

We have investigated the effects of 
RPV on the $b\to s\gamma\gamma$ transition amplitude, focusing 
on the two-photon B decays $B_s \to X_s\gamma\gamma$ 
and $B_s\to \gamma\gamma$.

We have calculated the complete RPV one-loop contribution to these decays
in the one-coupling scheme, i.e., assuming one contribution of a pair of RPV
couplings at a time. We 
found that the RPV effect is 
dominated by a single one-loop diagram which was named 
$\lambda$-{\it irreducible}. This diagram has a distinct topology 
which is irrelevant to $b \to s \gamma$ at one-loop. 
Therefore, since its effect on the one-photon decay $B \to X_s \gamma$ 
is negligible at one-loop, 
the $\lambda$-{\it irreducible} RPV effects that were found for 
the two photons decays
$B_s \to X_s\gamma\gamma$ 
and $B_s\to \gamma\gamma$ are uncorrelated to the decay 
$B \to X_s \gamma$. 

We found that the $\lambda$-{\it irreducible} RPV diagram with a $\tau$-loop 
gives 
$BR^{RPV}(B_s\to \gamma\gamma) \sim 16 \times 
BR^{SM}(B_s\to \gamma\gamma) \sim 
5 \times 10^{-6}$. 
In the case of $B\to X_s \gamma\gamma$ the enhancement to the decay width 
is less dramatic: $BR^{RPV}(B\to X_s \gamma\gamma) 
\sim 5 \times BR^{SM}(B\to X_s \gamma\gamma) \sim 
6 \times 10^{-7}$. 

We have also shown that, in spite of the rather marginal RPV effect 
on the decay width of $B \to X_s \gamma\gamma$, 
other observables may be used to 
disentangle the $\lambda$-{\it irreducible} RPV contribution.
In particular, we find that the energy distribution of the 
two photons is significantly different from the SM since 
it is sensitive to new threshold effects caused 
by the specific topology of the $\lambda$-{\it irreducible} diagram. 
Moreover, a forward-backward asymmetry with respect to the softer photon 
in $B \to X_s \gamma\gamma$ is appreciably different from its SM value
in the presence 
of the $\lambda$-{\it irreducible} RPV effect.    
Thus, these two observables may be used to provide an additional handle 
for discriminating different models, in particular RPV SUSY.

\begin{acknowledgments}
G.E. thanks the Technion President Fund for partial support.
\end{acknowledgments}

\end{document}